\documentclass[runningheads]{llncs}

\usepackage[T1]{fontenc}
\usepackage{graphicx}
\usepackage{amsmath,amssymb}
\usepackage{booktabs}
\usepackage{array}
\usepackage{float}
\usepackage{url}
\usepackage[normalem]{ulem}

\begin{document}
\raggedbottom

\title{The NPU-ASLP System for Music Restoration in the CCF-AATC 2026 Challenge}
\titlerunning{The NPU-ASLP System for the CCF-AATC 2026 Challenge Track 1}

\author{Zeyu Zhu\inst{1}$^{\dagger}$ \and
Yike Zhu\inst{1}$^{\dagger}$ \and
Xingchen Li\inst{1} \and
Chaoren Yu\inst{1} \and
Jingyi Wang\inst{1} \and
Yurun~Li\inst{1} \and
Jierui Guo\inst{1} \and
Chunjiang He\inst{1} \and
Lei Xie\inst{1}$^{*}$}
\authorrunning{Z. Zhu et al.}
\institute{Audio, Speech and Language Processing Group (ASLP@NPU), School of Computer Science,
Northwestern Polytechnical University, Xi'an, China\\
\url{https://www.npu-aslp.org}}

\maketitle
{
\renewcommand{\thefootnote}{\ensuremath{\dagger}}
\footnotetext{Equal contribution.}

\renewcommand{\thefootnote}{\ensuremath{*}}
\footnotetext{Corresponding author: lxie@nwpu.edu.cn.}
}

\begin{abstract}
This paper presents the NPU-ASLP music restoration system developed for Track~1
of the China Computer Federation (CCF) Advanced Audio Technology Competition
(AATC) 2026 Challenge.
The task aims to recover clean music from mobile-device-captured recordings
affected by coupled degradations from playback devices, acoustic propagation,
environmental noise and recording terminals.
System development is further constrained by limited and imperfectly aligned
real data pairs and a strict model-complexity budget.
In our system, we adopt TF-MoE as the backbone for its dual-axis modeling of cross-band and
temporal dependencies, whose Top-1 routing activates a single expert per token
to bound computation under the strict budget. We extend it with two
task-specific adaptations: a scale-aware encoder and a complex mapping decoder.
The former encodes normalized complex spectra jointly with per-band log-RMS to
preserve the band-energy cues discarded by band-wise normalization. The latter
directly estimates the clean complex spectrum rather than masking the input,
thereby repairing coupled degradations. To address the limited, imperfectly
aligned real recordings, we further build a challenge-calibrated
playback--propagation--recording simulator that generates 1,572.12~h of
sample-aligned data across 202 acoustic conditions. The final system contains 2.951M parameters and requires 82.659 GMACs
for a 10-s input. On the official blind test, it achieved -49.067~dB SI-SNR,
1.273 LSD, 2.540 FAD, and 3.495 ViSQOL. It ranked third in LSD, FAD, and
ViSQOL and received the third prize in the challenge.

\end{abstract}

\section{Introduction}

Music reproduced through amplification devices and re-recorded by mobile
terminals is common in user-generated and live-event content, yet the captured
signal can differ substantially from the source. The CCF-AATC 2026 Challenge Track~1
formalizes this problem in two settings: restoration after low-quality playback
and after reverberant multi-loudspeaker reproduction~\cite{aatc2026track1}.
Given a 16-kHz monaural recording, the goal is to recover
natural, clear music that is as close as possible to the clean reference.

This task is more than additive denoising. Loudspeaker coloration and
nonlinearity, distance attenuation, reverberation, environmental interference,
microphone response, and gain control are coupled in a single observation.
Several of these degradations act directly on band energy, substantially
weakening its absolute scale. Restoration is further complicated by music
itself, which combines broadband harmonics, fine overtones, percussive
transients, vocals, and diverse instruments; recovering it thus requires both
the fine complex-spectral structure and the energy relationships across bands.
Band-split representations suit this non-uniform frequency structure, but their
frame-wise, band-wise normalization discards the input amplitude scale and
weakens the cross-band level cues needed to compensate for the coupled degradations.

Learning this inverse mapping is also constrained by supervision and efficiency.
The officially provided pairs contain approximately 10~h of audio while
covering diverse devices, distances, scenes, and music types~\cite{aatc2026track1};
our analysis further found non-negligible temporal
offsets between some inputs and targets. Such offsets contaminate sample-aligned
spectral supervision. Meanwhile, the complex time-frequency variation demands
sufficient conditional capacity, yet model size and computation are bounded by
the challenge's strict parameter and computation budget.

Existing methods provide useful components but do not resolve these constraints
together. The official AudioFlow baseline performs iterative latent-space flow
matching, demonstrating both the feasibility of generative restoration and
robustness to misaligned training data, yet its 550.87M parameters and 4.49T
MACs are incompatible with the complexity requirement~\cite{audioflowbaseline}.
Band-Split RNN (BSRNN) shows the value of nonuniform subbands and alternating
cross-band and temporal modeling for music~\cite{bsrnn}, while Time-Frequency
Mixture-of-Experts (TF-MoE) adds
sparsely activated experts on both axes, trading model capacity against
activated computation~\cite{tfmoe}. Direct transfer, however, retains three task
mismatches: its band-wise normalization discards the amplitude scale needed
here, its masking head constrains the output to an input-dependent
multiplicative form, and it does not supply the aligned supervision required by
the limited, misaligned real data pairs. The gap is therefore not that existing
methods generally underperform, but how to jointly recover the coupled
degradations, preserve the input amplitude-scale cues, and obtain large-scale,
strictly aligned supervision matched to the challenge conditions, all within a
limited parameter budget.

To address these issues, we combine direct complex mapping for coupled
degradation repair, scale-aware encoding for band-energy preservation, dual-axis
Top-1 TF-MoE for compact time-frequency modeling, and challenge-calibrated
simulation that produces 1,572.12~h of source-disjoint aligned pairs over
202 conditions for aligned supervision. Eventually, the resulting
2.951M-parameter system received the third prize, with official blind-test
scores of $-49.067$~dB SI-SNR, 1.273 LSD, 2.540 FAD, and 3.495
ViSQOL~\cite{aatc2026track1}.

\section{Method}

\subsection{Scale-Aware TF-MoE Model}
\subsubsection{Backbone Architecture}
We adopt the Time-Frequency Mixture-of-Experts (TF-MoE)~\cite{tfmoe} as the
restoration backbone with two key modifications: (i) its band-wise
normalization is replaced with a scale-aware band encoder, and (ii) its mask decoding module is replaced with a complex mapping decoder that directly estimates the target complex spectrum.

\begin{figure}[htbp]
\centering
\includegraphics[width=\linewidth]{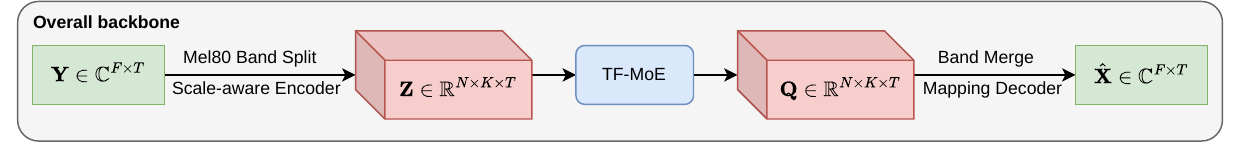}
\caption{Overview of the restoration backbone. The input spectrogram is
extracted using a 32 ms Hanning window with an 8 ms shift, and split into
$K=80$ Mel-style sub-bands; the model uses a hidden dimension of $N=32$ and
$R=6$ stacked TF-MoE blocks.}
\label{fig:overview}
\end{figure}

As shown in Fig.~\ref{fig:overview}, the backbone takes the complex spectrum
$\mathbf{Y}\in\mathbb{C}^{F\times T}$ of the recorded waveform as input, where $F$ is
the number of frequency bins and $T$ denotes the number of time frames. The
band splitting module partitions the $F$ frequency bins into $K$ Mel-style
bands, and the scale-aware band encoder encodes the frequency bins in each band
into an $N$-dimensional feature, resulting in a
three-dimensional representation $\mathbf{Z}\in\mathbb{R}^{N\times K\times T}$.
Six stacked TF-MoE blocks then process $\mathbf{Z}$ along the frequency and time
axes, and the mapping decoder merges the subband features and directly
estimates the target complex spectrum $\hat{\mathbf{X}}\in\mathbb{C}^{F\times T}$, from
which the restored waveform is reconstructed by the inverse STFT.

\subsubsection{Scale-Aware Band Encoding}
\begin{figure}[htbp]
    \centering
    \includegraphics[width=\linewidth]{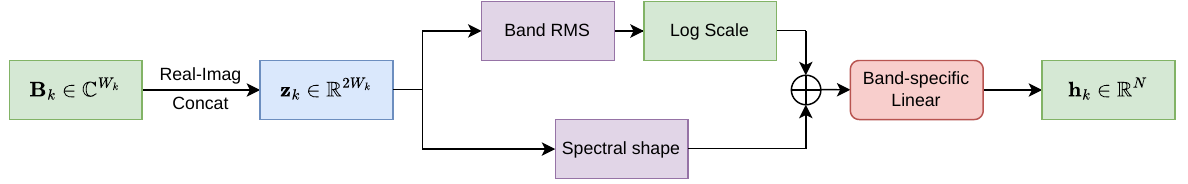}
    \caption{Scale-aware band encoding. One branch divides the complex spectrum by
    its per-band RMS to preserve phase and within-band structure, while the other
    branch computes the logarithmically compressed RMS; the two are concatenated
    and fed to a band-specific linear projection.}
    \label{fig:scale-aware-detail}
\end{figure}

To retain amplitude cues removed by conventional subband normalization, we
extract the absolute amplitude scale as an explicit feature. For band $b$ in frame $t$,
the real and imaginary coefficients are concatenated as
$z_{t,b}=[\operatorname{Re}(Y_{t,b});\operatorname{Im}(Y_{t,b})]
\in\mathbb{R}^{2W_b}$, where $W_b$ is the band width. We then compute
\begin{equation}
\begin{aligned}
r_{t,b} &= \sqrt{\frac{1}{2W_b}\lVert z_{t,b}\rVert_2^2+\epsilon},
&\mathbf d_{t,b} &= \frac{z_{t,b}}{r_{t,b}},\\
h_{t,b}^{(0)} &= \phi_b\!\left([\mathbf d_{t,b};\log(r_{t,b}+\epsilon)]\right),
\end{aligned}
\label{eq:scale}
\end{equation}
where $\epsilon=10^{-8}$ and $\phi_b$ projects to $N$ dimensions. As shown in
Fig.~\ref{fig:scale-aware-detail}, we first compute the per-band RMS $r_{t,b}$.
One branch divides the complex spectrum by this RMS, yielding the unit-RMS
direction $\mathbf d_{t,b}$ that preserves the phase and the relative spectral
relation within the band while constraining the numerical range of the input
feature; the other branch computes the logarithmically compressed RMS, which
tells the model how strong this band actually is. The two branches are then
concatenated and encoded jointly, so the scale is explicitly retained alongside
the normalization.

\begin{figure}[htbp]
    \centering
    \includegraphics[width=\linewidth]{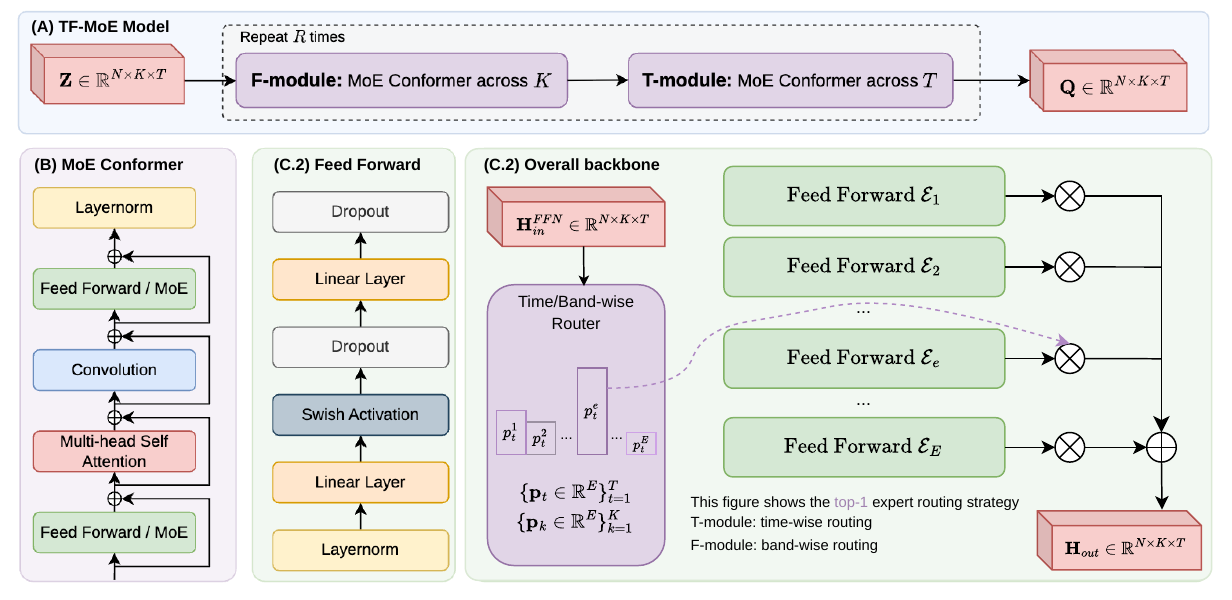}
    \caption{Dual-axis TF-MoE architecture. This diagram is adapted
    from~\cite{tfmoe} and revised to reflect the configuration used in this work.}
    \label{fig:dual-axis}
\end{figure}

\subsubsection{Dual-Axis Sparse Modeling}
Six TF-MoE blocks sequentially model cross-band structure and temporal dynamics
with frequency-axis (F) and time-axis (T) modules, respectively
(Fig.~\ref{fig:dual-axis}). Each module follows a Conformer-style design with
four-head self-attention, a kernel-31 convolution, and two MoE feed-forward
layers~\cite{conformer,tfmoe}. Each MoE contains $E=12$ two-layer feed-forward
experts with expansion factor four. For token $h_i$, the Top-1 router activates the highest-probability expert:
\begin{equation}
p_i=\operatorname{softmax}(W_g h_i),\qquad
e_i^*=\arg\max_e p_{i,e},\qquad
\operatorname{MoE}(h_i)=p_{i,e_i^*}\mathcal{E}_{e_i^*}(h_i).
\label{eq:moe}
\end{equation}
Although the hard choice $e_i^*$ is non-differentiable, the selected gate value $p_{i,e_i^*}$ carries reconstruction gradients to the router.

\subsubsection{Complex Spectral Mapping Decoder}
\begin{figure}[htbp]
    \centering
    \includegraphics[width=\linewidth]{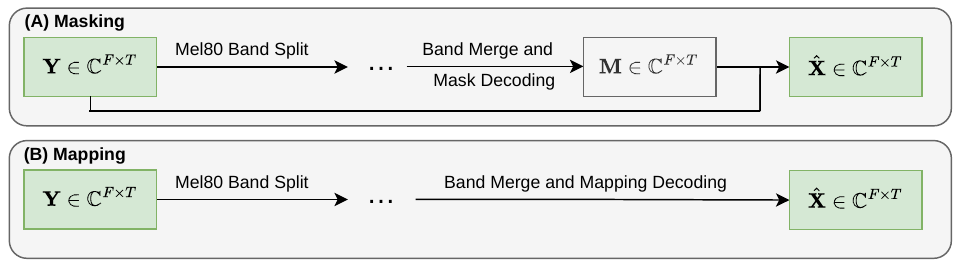}
    \caption{Complex masking versus the proposed direct mapping decoder.}
    \label{fig:map-detail}
\end{figure}

As shown in Fig.~\ref{fig:map-detail}, the decoder directly predicts the real and imaginary parts of the target
complex spectrum instead of estimating a mask to multiply with the input.
This remains conditioned on the degraded input $Y$ but no longer
constrains each estimate to be a pointwise product of $Y$, which is more
flexible for components that have been strongly attenuated or nonlinearly
distorted.

\subsubsection{Training Objective}
Our training objective combines spectral, waveform, perceptual, and
expert-balancing terms. The reconstruction objective contains three losses. A
multi-resolution STFT loss $\mathcal{L}_{\mathrm{MRSTFT}}$ combines
compressed-magnitude and complex-domain MSE terms. A time-domain MAE loss is
denoted by $\mathcal{L}_{\mathrm{time}}$, while
$\mathcal{L}_{\mathrm{VGGish}}$ is computed using a frozen VGGish
model~\cite{vggish}.

To encourage balanced expert utilization and prevent router collapse, we further adopt a Switch-style load-balancing loss~\cite{switch}. For a router with $I$ input tokens and $E=12$ experts, the soft importance and hard load of expert $e$ are defined as
\begin{equation}
P_e = \frac{1}{I}\sum_{i=1}^{I}p_{i,e},
\qquad
f_e = \frac{1}{I}\sum_{i=1}^{I}\mathbf{1}\!\left[e_i^*=e\right],
\end{equation}
where $p_{i,e}$ is the routing probability of token $i$ for expert $e$, and $e_i^*$ denotes the selected expert for token $i$. The router-level balancing loss is
\begin{equation}
\ell_{\mathrm{bal}}
= E\sum_{e=1}^{E}P_e\operatorname{sg}(f_e),
\label{eq:bal-router}
\end{equation}
where $\operatorname{sg}(\cdot)$ denotes stop-gradient. Hence, gradients are propagated through the differentiable importance $P_e$, while the discrete load $f_e$ is treated as constant.

Each TF-MoE block contains two MoE feed-forward layers along each of the frequency and time axes, yielding four router losses per block. We sum these losses within each block and average them across the $L=6$ blocks:
\begin{equation}
\mathcal{L}_{\mathrm{bal}}
= \frac{1}{L}\sum_{\ell=1}^{L}
\sum_{a\in\{\mathrm{F},\mathrm{T}\}}\sum_{q=1}^{2}
\ell_{\mathrm{bal}}^{(\ell,a,q)},
\qquad L=6.
\label{eq:bal-total}
\end{equation}

The total loss is
\begin{equation}
\mathcal{L}
= \mathcal{L}_{\mathrm{MRSTFT}}
+ 0.2\,\mathcal{L}_{\mathrm{time}}
+ \lambda_{\mathrm{perc}}(s)\,\mathcal{L}_{\mathrm{VGGish}}
+ 0.001\,\mathcal{L}_{\mathrm{bal}}.
\label{eq:loss}
\end{equation}
The perceptual weight is linearly warmed up to $0.05$ over the first $2{,}000$ optimizer steps:
\begin{equation}
\lambda_{\mathrm{perc}}(s)
= 0.05\min\!\left(\frac{s}{2{,}000},1\right),
\qquad s\geq 0,
\label{eq:perc-weight}
\end{equation}
where $s$ denotes the global optimizer step.

\subsection{Challenge-calibrated Simulation Pipeline}

The official data pairs~\cite{aatc2026track1} are useful for profiling the target domain, but
are limited and not consistently aligned, which may lead to unstable training
of discriminative models. Specifically, the cross-correlation analysis revealed a
mean absolute delay of 89~ms, and delays above 20~ms comprise 80\% of the pairs. We
therefore use the real recordings to estimate scene, device, distance, and level
distributions rather than as the sole source of sample-aligned supervision.

\subsubsection{Source Material and Condition Design}
Table~\ref{tab:data} summarizes the source pools. High-quality music is
collected from the web and screened by automatic tagging and listening. The RIR
bank is generated with Pyroomacoustics~\cite{pyroomacoustics}, while the noise pool, all listed in Table~\ref{tab:noise}, includes public corpora and internally collected web audio. All sources
are converted to 16-kHz mono PCM. In total, we
sample 1,000 episodes for each of 202 scene-device-quality-distance
conditions, generating 202,000 sample-aligned pairs,
adding up to 1,572.121~h.

\begin{table}[H]
\caption{Source material and generated simulation data.}
\label{tab:data}
\centering
\small
\setlength{\tabcolsep}{3.5pt}
\begin{tabular}{lrl}
\toprule
Source & Scale & Role \\
\midrule
Official pairs & $\sim$10~h & profiling/real validation \\
Selected music & 180,436 entries & clean targets \\
Noise & 821,144 entries & ambience/events \\
RIR bank & 50,000 responses & propagation \\
Generated & 202,000 episodes; 1,572.121~h & aligned pairs \\
\bottomrule
\end{tabular}
\end{table}

\begin{table}[H]
\caption{Noise datasets used by the simulator.}
\label{tab:noise}
\centering
\footnotesize
\setlength{\tabcolsep}{12pt}
\renewcommand{\arraystretch}{1.08}
\begin{tabular}{@{}ll@{}}
\toprule
\multicolumn{2}{c}{Dataset} \\
\midrule
DNS Challenge 3/5~\cite{dnschallenge,dnschallenge5} & MUSAN noise~\cite{musan} \\
TAU Urban Acoustic Scenes~\cite{tau2020} & FSD50K~\cite{fsd50k} \\
DEMAND~\cite{demand} & UrbanSound8K~\cite{urbansound8k} \\
Datumo Background Noise~\cite{datumo} & ESC-50~\cite{esc50} \\
ASPED v.b~\cite{aspedvb} & SONYC-UST~\cite{sonycust} \\
Web-collected audio & \\
\bottomrule
\end{tabular}
\end{table}

\subsubsection{Physical Degradation Pipeline}
\begin{figure}[htbp]
    \centering
    \includegraphics[width=\linewidth]{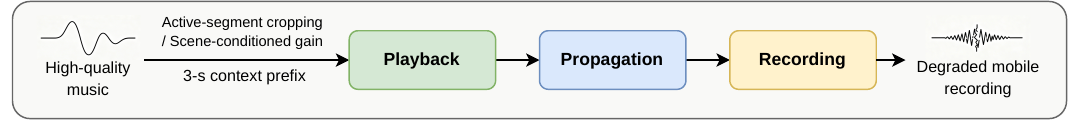}
    \caption{Overview of the challenge-calibrated simulation pipeline. Active
    high-quality music is level-conditioned with a 3-s context prefix, degraded by
    playback and acoustic propagation, and finally processed by a mobile-recording
    front end.}
    \label{fig:sim-overview}
\end{figure}

Each episode follows the physical recording order shown in Fig.~\ref{fig:sim-overview}. \textbf{Pre-processing} prepares an active music excerpt, matches the official clean RMS profile, and retains a 3-s real context prefix. \textbf{Playback} simulates the device degradation as the music leaves the player, and \textbf{Propagation} adds the spatial and environmental degradation. Finally, \textbf{Recording} applies the mobile-device recording chain. Details of these stages are presented below.

\begin{figure}[htbp]
\centering
\includegraphics[width=\linewidth]{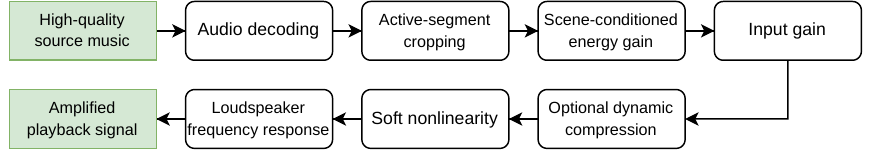}
\caption{Pre-processing and playback simulation. High-quality music is decoded,
cropped to an active segment, and adjusted with a scene-conditioned energy
gain. The playback model then applies input gain, optional dynamic compression,
soft nonlinearity, and loudspeaker frequency response to produce the amplified
playback signal.}
\label{fig:sim-preplay}
\end{figure}

\paragraph{Pre-processing and Playback}
As depicted in Fig.~\ref{fig:sim-preplay}, after decoding the high-quality source,
we select an active 20--36-s excerpt and apply a scene-conditioned energy gain
sampled to match the official clean RMS profile. A real 3-s prefix from the same music is retained as processing context and discarded from the final output, allowing reverberation,
compression, and AGC states to settle before the supervised segment. The
playback stage then randomly applies input gain, optional dynamic compression,
soft nonlinearity, and loudspeaker frequency response to produce the amplified signal.

\begin{figure}[htbp]
\centering
\includegraphics[width=0.82\linewidth]{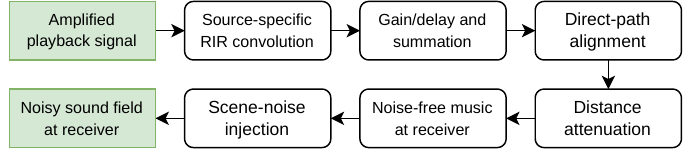}
\caption{Propagation simulation. The propagation stage builds a single- or
multi-loudspeaker sound field, aligns the direct path, applies distance
attenuation, and injects scene noise.}
\label{fig:sim-propagation}
\end{figure}

\paragraph{Propagation}
As illustrated in Fig.~\ref{fig:sim-propagation}, starting from the amplified signal, we convolve each loudspeaker channel with a
source-specific RIR, apply an independent gain and a small delay, and sum the
channels to form the spatial sound field. Conventional devices predominantly
use one or two sources, whereas loudspeaker arrays use 6, 12, or 18. We then
locate the direct path and compensate for the convolution delay so that the
three supervision signals share the same starting sample. A scene-dependent
distance gain covers the official 1--50-m conditions~\cite{aatc2026track1} and produces the
noise-free music at the receiver. Finally, continuous background noise, sparse
events, or their mixture is injected at an SNR sampled from $-15$ to 20~dB.

\begin{figure}[h]
\centering
\includegraphics[width=\linewidth]{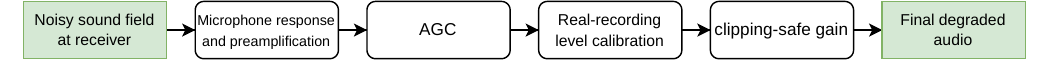}
\caption{Recording simulation. The recording stage models the microphone/front
end, shared AGC, level calibration, and clipping-safe gain.}
\label{fig:sim-recording}
\end{figure}

\paragraph{Recording} Figure~\ref{fig:sim-recording} shows the mobile-device
recording chain. The signal first undergoes microphone response filtering and
preamplification, followed by an AGC stage that emulates the phone's automatic
gain control. It is
subsequently calibrated to the official real-recording level distribution and
scaled by a linear safety gain that prevents clipping, yielding the final
degraded audio.

To keep the simulated data close to the real recordings, all simulator
parameters are selected and tuned on the official data, and are further
adjusted according to the scene and the quality level of each playback device.

\section{Experiments}

\subsection{Training and Evaluation Setup}
We train from random initialization with crop-based updates, and select the
checkpoint on both simulated validation episodes and the official development
recordings, so that model selection reflects the simulated and the real domain.
Inference follows a chunked scheme with overlap-add, and optimization uses
AdamW~\cite{adamw}. Table~\ref{tab:config} lists the complete configuration.

\begin{table}[htbp]
\caption{System and training configuration.}
\label{tab:config}
\centering
\small
\begin{tabular}{@{}p{0.24\linewidth}p{0.66\linewidth}@{}}
\toprule
Item & Configuration \\
\midrule
Input & 16-kHz mono waveform \\
STFT & Hann window; FFT size/window/hop = 512/512/128 samples \\
Initialization & random, seed 42 \\
Training data & 200,000 simulated episodes; random aligned 4-s crops \\
Validation & start/middle/end 10-s crops of the 25 official development recordings and 500 fixed simulated episodes \\
Optimizer & AdamW, LR $3\times10^{-4}$, weight decay 0.01 \\
Schedule & 2,000-step linear warm-up, cosine decay \\
Batch & 8 GPUs $\times$ 12 = global batch 96 \\
Inference & 10-s chunks; 5-s hop; squared sine/cosine overlap-add \\
\bottomrule
\end{tabular}
\end{table}

We report SI-SNR~\cite{sisnr}, LSD~\cite{lsd}, FAD~\cite{fad}, and
ViSQOL~\cite{visqol};
the first and last are higher-is-better. Development results cover the full
180~s of all 25 real recordings. The official blind set is reported separately
and was never used for training, tuning, selection, or simulator calibration.
Complexity includes the STFT and inverse STFT for one 10-s waveform, follows the
official $1$~MAC$=2$~FLOPs convention~\cite{aatc2026track1}, and excludes window
overlap.

\subsection{Development Dataset Results}
Table~\ref{tab:dev} compares the degraded input, our local rerun of the official baseline, and the submitted system under a common full-track evaluator. The results reveal that the system achieves a better quality--complexity trade-off, surpassing the baseline on all four metrics at a much lower parameter and computation cost.
Specifically, compared with the baseline, our system improves SI-SNR by 10.73~dB, LSD by 25.6\%, FAD by 44.8\%, and ViSQOL by 0.105, while using roughly $1/187$ of the parameters and $1/54$ of the MACs.
The gains are most evident in LSD and FAD, showing that the model effectively restores the spectral structure of the music and substantially narrows the distribution gap between the output and high-quality music; the ViSQOL of 3.475 indicates a clear perceptual improvement as well.

\begin{table}[htbp]
\caption{Full-track results on 25 official real development recordings. The
baseline row is our local rerun using the common evaluator. Boldface marks the
best value in each column, and underlining marks the second-best value.}
\label{tab:dev}
\centering
\small
\setlength{\tabcolsep}{4.0pt}
\begin{tabular}{lrrrrrr}
\toprule
System & SI-SNR$\uparrow$ & LSD$\downarrow$ & FAD$\downarrow$ & ViSQOL$\uparrow$ & Params/M$\downarrow$ & GMACs$\downarrow$ \\
\midrule
Input & \underline{-46.659} & 1.887 & 19.127 & 3.186 & -- & -- \\
Baseline & -56.170 & \underline{1.700} & \underline{6.430} & \underline{3.370} & \underline{550.87} & \underline{4490} \\
\textbf{Ours} & \textbf{-45.440} & \textbf{1.265} & \textbf{3.549} & \textbf{3.475} & \textbf{2.951} & \textbf{82.659} \\
\bottomrule
\end{tabular}
\end{table}

\subsection{Official Blind Test Results}
Table~\ref{tab:blind} reports the released leaderboard metrics~\cite{audioflowbaseline} for the six
finalist teams in CCF-AATC 2026 Track 1. Our team (T010) received the third prize
and ranked third in LSD, FAD, and
ViSQOL. With only 2.951M parameters, our system achieves competitive spectral and perceptual results. Although it is
not the smallest entry, it still belongs to the highest tier of the official evaluation in terms of model size (fewer than 5M parameters).

The SI-SNR scores are uniformly low across the six systems, which may stem
from the sensitivity of SI-SNR to sample-level alignment; the metric therefore
has limited reference value for this task.

\begin{table}[t]
    \caption{Official blind-test results of the finalist teams in CCF-AATC 2026 Track 1.
    Boldface marks the best value in
    each column among the six public systems, underlining marks the second-best,
    and a wavy underline the third-best; our system ID is also shown in
    boldface.}
    \label{tab:blind}
    \centering
    \small
    \setlength{\tabcolsep}{3.3pt}
    \begin{tabular}{lrrrrrr}
    \toprule
    ID & SI-SNR$\uparrow$ & LSD$\downarrow$ & FAD$\downarrow$ & ViSQOL$\uparrow$ & Params/M$\downarrow$ & GMACs$\downarrow$ \\
    \midrule
    T106 & -47.213 & \textbf{1.102} & \textbf{0.540} & \textbf{3.786} & 364.805 & 614.993 \\
    T088 & \underline{-40.530} & \underline{1.269} & \underline{2.056} & 3.446 & \uwave{2.915} & \uwave{56.437} \\
    T034 & \uwave{-43.445} & 1.373 & 2.651 & 3.428 & 4.768 & 1592.067 \\
    \textbf{T010 (ours)} & -49.067 & \uwave{1.273} & \uwave{2.540} & \uwave{3.495} & 2.951 & 82.659 \\
    T026 & \textbf{-39.765} & 1.448 & 3.720 & 3.406 & \underline{1.145} & \textbf{0.933} \\
    T006 & -44.852 & 1.309 & 14.950 & \underline{3.528} & \textbf{0.637} & \underline{11.193} \\
    \bottomrule
    \end{tabular}
\end{table}

\subsection{Ablation Studies}
To validate the contribution of each component, we perform separate ablation experiments on scale-aware encoding, prediction targets, and loss profiles. Although the training configuration differs slightly from that of the submitted system, all ablation variants use identical settings except for the factor listed in Table~\ref{tab:ablation}; in each comparison, only that factor is varied.

\begin{table}[htbp]
    \caption{Results of the ablation studies. Boldface marks the best value
    within each factor group.}
    \label{tab:ablation}
    \centering
    \small
    \setlength{\tabcolsep}{4.5pt}
    \begin{tabular}{llrrrr}
    \toprule
    Factor & Variant & SI-SNR$\uparrow$ & LSD$\downarrow$ & FAD$\downarrow$ & ViSQOL$\uparrow$ \\
    \midrule
    Encoder & Legacy & -46.1614 & 1.3400 & 5.5563 & 3.3804 \\
     & Scale-aware & \textbf{-43.2666} & \textbf{1.3370} & \textbf{5.2871} & \textbf{3.3883} \\
    \midrule
    Output & Hybrid & -43.9899 & 1.3715 & \textbf{4.8409} & 3.3615 \\
     & Mask & -43.6849 & 1.4453 & 6.0527 & 3.3096 \\
     & Mapping & \textbf{-43.2666} & \textbf{1.3370} & 5.2871 & \textbf{3.3883} \\
    \midrule
    Loss & Legacy & -49.1489 & 1.8318 & 21.5338 & 3.1569 \\
     & Competition & \textbf{-43.9899} & \textbf{1.3715} & \textbf{4.8409} & \textbf{3.3615} \\
    \bottomrule
    \end{tabular}
\end{table}

\subsubsection{Scale-Aware Encoder}
We first replace the legacy encoder with the scale-aware encoder while keeping complex mapping and the competition loss profile fixed. The scale-aware encoder outperforms the legacy encoder on all four metrics: SI-SNR increases by
2.895~dB and ViSQOL by 0.008, while LSD and FAD decrease by 0.003 and 0.269, respectively. This validates the effectiveness of scale-aware encoding.

\subsubsection{Prediction Target}
We next compare hybrid estimation, complex masking, and direct complex mapping with the scale-aware encoder and competition loss profile fixed. The three heads produce $\hat S=(Y+U)\odot M$, $\hat S=Y\odot M$, and $\hat S=U$, respectively, where $U$ denotes an unconstrained complex estimate. Mapping achieves the best SI-SNR, LSD, and ViSQOL, while hybrid estimation achieves the lowest FAD. Relative to hybrid estimation, mapping improves SI-SNR by 0.723~dB, LSD by 0.035, and ViSQOL by 0.027, at the cost of a 0.446 increase in FAD. It is therefore selected for its overall multi-metric trade-off rather than dominance on every metric; masking does not outperform mapping on any of the four metrics at the matched checkpoint. Additionally, at each time--frequency bin, the hybrid head predicts both an additive complex component and a complex mask, corresponding to four real-valued outputs, whereas the masking and mapping heads each predict only two. Consequently, the wider output projection of the hybrid head incurs slightly more decoder parameters and MACs.

\subsubsection{Loss Profile}
Finally, we compare the legacy and competition-oriented objectives with the scale-aware encoder and hybrid head fixed. The competition loss profile improves all four metrics, with gains of 5.159~dB in SI-SNR and 0.205 in ViSQOL, together with reductions of 0.460 in LSD and 16.693 in FAD. This comparison changes the spectral loss formulation, removes the scale-invariant SI-SNR term, and adds waveform $L_1$ and frozen VGGish feature terms jointly. These results demonstrate that the competition-oriented loss profile provides a substantially more effective training objective for the target restoration task.

\section{Conclusion}

This paper presented a system for restoring music affected by coupled playback,
propagation, and mobile-recording degradations. The system combines
challenge-calibrated simulation for aligned supervision with a scale-aware
encoder, direct complex mapping, and a dual-axis Top-1 TF-MoE backbone.
The ablations validate scale-aware encoding and the competition-oriented
objective. Direct complex mapping provides the best overall trade-off among the
evaluated prediction targets. With 2.951M
parameters and 82.659 GMACs per 10-s input, the submitted system received the
third prize and ranked third in LSD, FAD, and ViSQOL on the official blind test,
achieving 1.273, 2.540, and 3.495, respectively. The weaker SI-SNR result and
the remaining computation cost define the main limitations. Future work will
target sample-alignment robustness and lower computation.

\bibliographystyle{splncs04}
\bibliography{tfmoe_paper}

\begin{thebibliography}{10}
\providecommand{\url}[1]{\texttt{#1}}
\providecommand{\urlprefix}{URL }
\providecommand{\doi}[1]{https://doi.org/#1}

\bibitem{sonycust}
Cartwright, M., Mendez~Mendez, A.E., Cramer, J., et~al.: {SONYC} urban sound tagging ({SONYC-UST}): A multilabel dataset from an urban acoustic sensor network. In: Proc. DCASE Workshop. pp. 35--39 (2019)

\bibitem{aatc2026track1}
{CCF Advanced Audio Technology Competition Organizing Committee}: {CCF-AATC 2026 Track 1: Music Restoration after Amplified Playback}. \url{https://ccf-aatc.org.cn/2026} (2026), accessed 15 September 2026

\bibitem{visqol}
Chinen, M., Lim, F.S.C., Skoglund, J., et~al.: {ViSQOL v3}: An open source production ready objective speech and audio metric. In: Proc. QoMEX. pp.~1--6 (2020)

\bibitem{datumo}
{Datumo}: Background noise dataset. \url{https://open.datumo.com/en/?page_id=5764}, last accessed 19 September 2026

\bibitem{dnschallenge5}
Dubey, H., Aazami, A., Gopal, V., Naderi, B., Braun, S., Cutler, R., Ju, A., Zohourian, M., Tang, M., Golestaneh, M., Aichner, R.: {ICASSP 2023} deep noise suppression challenge. IEEE Open Journal of Signal Processing  \textbf{5},  725--737 (2024). \doi{10.1109/OJSP.2024.3378602}

\bibitem{switch}
Fedus, W., Zoph, B., Shazeer, N.: Switch transformers: Scaling to trillion parameter models with simple and efficient sparsity. Journal of Machine Learning Research  \textbf{23}(120),  1--39 (2022)

\bibitem{fsd50k}
Fonseca, E., Favory, X., Pons, J., et~al.: {FSD50K}: An open dataset of human-labeled sound events. IEEE/ACM Transactions on Audio, Speech, and Language Processing  \textbf{30},  829--852 (2022)

\bibitem{lsd}
Gray, A.H., Markel, J.D.: Distance measures for speech processing. IEEE Transactions on Acoustics, Speech, and Signal Processing  \textbf{24}(5),  380--391 (1976). \doi{10.1109/TASSP.1976.1162849}

\bibitem{conformer}
Gulati, A., Qin, J., Chiu, C.C., et~al.: Conformer: Convolution-augmented transformer for speech recognition. In: Proc. Interspeech. pp. 5036--5040 (2020)

\bibitem{tau2020}
Heittola, T., Mesaros, A., Virtanen, T.: Acoustic scene classification in {DCASE 2020} challenge: Generalization across devices and low complexity solutions. In: Proc. DCASE Workshop. pp. 56--60 (2020)

\bibitem{vggish}
Hershey, S., Chaudhuri, S., Ellis, D.P.W., et~al.: {CNN} architectures for large-scale audio classification. In: Proc. IEEE ICASSP. pp. 131--135 (2017)

\bibitem{tfmoe}
Hu, Q., Li, C., Zhang, W., Liu, S., Lu, Y., Qian, Y.: {TF-MoE}: Time-frequency mixture-of-experts for efficient speech separation. In: Proc. Interspeech. pp. 3172--3177 (2026). \doi{10.21437/Interspeech.2026-1307}

\bibitem{fad}
Kilgour, K., Zuluaga, M., Roblek, D., Sharifi, M.: Fr\'{e}chet audio distance: A reference-free metric for evaluating music enhancement algorithms. In: Proc. Interspeech. pp. 2350--2354 (2019)

\bibitem{aspedvb}
Kim, Y., Han, C., Sarode, A., Posner, N., Guhathakurta, S., Lerch, A.: Audio-based pedestrian detection in the presence of vehicular noise. In: Proc. DCASE Workshop. pp. 60--64 (2025). \doi{10.5281/zenodo.17251589}

\bibitem{sisnr}
Le~Roux, J., Wisdom, S., Erdogan, H., Hershey, J.R.: {SDR}---half-baked or well done? In: Proc. IEEE ICASSP. pp. 626--630 (2019)

\bibitem{audioflowbaseline}
Liu, F.: {AudioFlow}: Audio generation with flow matching for the {CCF-AATC 2026 Track 1}. GitHub repository, \url{https://github.com/qiuqiangkong/audioflow/tree/CCF_AATC_2026_1} (2026), branch: \texttt{CCF\_AATC\_2026\_1}; accessed 15 September 2026

\bibitem{adamw}
Loshchilov, I., Hutter, F.: Decoupled weight decay regularization. In: Proc. ICLR (2019)

\bibitem{bsrnn}
Luo, Y., Yu, J.: Music source separation with band-split {RNN}. IEEE/ACM Transactions on Audio, Speech, and Language Processing  \textbf{31},  1893--1901 (2023)

\bibitem{esc50}
Piczak, K.J.: {ESC}: Dataset for environmental sound classification. In: Proc. ACM Multimedia. pp. 1015--1018 (2015)

\bibitem{dnschallenge}
Reddy, C.K.A., Dubey, H., Koishida, K., et~al.: {INTERSPEECH 2021} deep noise suppression challenge. In: Proc. Interspeech. pp. 2796--2800 (2021)

\bibitem{urbansound8k}
Salamon, J., Jacoby, C., Bello, J.P.: A dataset and taxonomy for urban sound research. In: Proc. ACM Multimedia. pp. 1041--1044 (2014)

\bibitem{pyroomacoustics}
Scheibler, R., Bezzam, E., Dokmani\'{c}, I.: Pyroomacoustics: A python package for audio room simulation and array processing algorithms. In: Proc. IEEE ICASSP. pp. 351--355 (2018)

\bibitem{musan}
Snyder, D., Chen, G., Povey, D.: {MUSAN}: A music, speech, and noise corpus. arXiv preprint arXiv:1510.08484 (2015)

\bibitem{demand}
Thiemann, J., Ito, N., Vincent, E.: The diverse environments multi-channel acoustic noise database ({DEMAND}): A database of multichannel environmental noise recordings. Proceedings of Meetings on Acoustics  \textbf{19}(1),  035081 (2013). \doi{10.1121/1.4799597}

\end{thebibliography}

\end{document}